\documentclass[fleqn,usenatbib]{mnras}

\usepackage{newtxtext,newtxmath}

\usepackage[T1]{fontenc}

\DeclareRobustCommand{\VAN}[3]{#2}
\let\VANthebibliography\thebibliography
\def\thebibliography{\DeclareRobustCommand{\VAN}[3]{##3}\VANthebibliography}

\usepackage{graphicx}	
\usepackage{amsmath}	
\usepackage{placeins,float,times,pifont}
\usepackage{multirow}
\usepackage[dvipsnames]{xcolor}

\newcommand{\bq}{\begin{eqnarray}}
\newcommand{\eq}{\end{eqnarray}}

\title[Degeneracies between baryons and dark matter]{Degeneracies between baryons and dark matter: the challenge of constraining the nature of dark matter with JWST}

\author[Khimey, Bose \& Tacchella]{
Diana Khimey,$^{1}$\thanks{E-mail: dkhimey@college.harvard.edu}
Sownak Bose,$^{1}$\thanks{E-mail: sownak.bose@cfa.harvard.edu}
Sandro Tacchella$^{1}$\thanks{E-mail: sandro.tacchella@cfa.harvard.edu}
\\
$^{1}$Center for Astrophysics $\vert$ Harvard \& Smithsonian, 60 Garden St, Cambridge, MA 02138, USA
}

\date{Accepted XXX. Received YYY; in original form ZZZ}

\pubyear{2020}

\begin{document}
\label{firstpage}
\pagerange{\pageref{firstpage}--\pageref{lastpage}}
\maketitle

\begin{abstract}
The James Webb Space Telescope (JWST) will revolutionise our understanding of early galaxy formation, and could potentially set stringent constraints on the nature of dark matter. We use a semi-empirical model of galaxy formation to investigate the extent to which uncertainties in the implementation of baryonic physics may be degenerate with the predictions of two different models of dark matter -- Cold Dark Matter (CDM) and a 7 keV sterile neutrino, which behaves as Warm Dark Matter (WDM). Our models are calibrated to the observed UV luminosity function at $z=4$ using two separate dust attenuation prescriptions, which manifest as high and low star formation efficiency in low mass haloes. We find that while at fixed star formation efficiency, $\varepsilon$, there are marked differences in the abundance of faint galaxies in the two dark matter models at high-$z$, these differences are mimicked easily by varying $\varepsilon$ in the same dark matter model. We find that a high $\varepsilon$ WDM and a low $\varepsilon$ CDM model -- which provide equally good fits to the $z=4$ UV luminosity function -- exhibit nearly identical evolution in the cosmic stellar mass and star formation rate densities. We show that differences in the star formation rate at fixed stellar mass are larger for variations in $\varepsilon$ in a {\it given} dark matter model than they are {\it between} dark matter models; however, the scatter in star formation rates is larger between the two models than they are when varying $\varepsilon$. Our results suggest that JWST will likely be more informative in constraining baryonic processes operating in high-$z$ galaxies than it will be in constraining the nature of dark matter. 

\end{abstract}




\section{Introduction}
The current paradigm of galaxy formation relies on the assumption that galaxies form out of the condensation of gas within dark matter haloes \citep{White_1978}. These haloes themselves form as particles of dark matter clump together under gravitational instability. Dark matter, then, clearly plays a fundamental role in the galaxy formation process, however its detailed nature is still unknown because dark matter particles have not yet been  detected directly. 

The Lambda Cold Dark Matter ($\Lambda$CDM) model \citep{Blumenthal_1984} is the current standard model of halo structure formation, and has been remarkably successful at detailing the distribution of matter in the Universe at large-scales, as well as successfully predicting the properties of the galactic population across cosmic time. That said, the model has faced challenges when $N$-body simulations have been used to probe the small-scale structure of dark matter haloes, initially showing discrepancies with existing observations of low-mass dwarf galaxies. Perhaps the most prominent amongst these are the so-called `Missing Satellites' \citep[][]{Klypin_1999,Moore_1999}, the `Too Big to Fail' \citep{BoylanKolchin_2011}, the `Cusp-core' \citep{Flores_1994,deBlok_2001} and, more recently, the so-called `Planes of Satellites' \citep{Ibata_2014,Pawlowski_2014} problems. A comprehensive review of these issues is presented in \cite{Bullock_2017}. While alternative candidates to CDM have often been proposed as a means to bypass these issues, the development of more sophisticated treatment of baryonic physics -- or, more generally, in the comparison of cosmological simulations with observations -- has been largely successful at resolving discrepancies within a CDM framework \citep[e.g.][]{Kauffmann_1993,Pontzen_2012,Brooks_2013,DiCintio_2014,Cautun_2015,Sawala_2016,Read_2017,Kim_2018,Genina_2018}. In particular, modern generations of hydrodynamical simulations show that the effect of stellar and radiative feedback in galaxies suppress star formation, with the net result that not all dark matter haloes end up hosting a galaxy, providing a natural solution to the missing satellites problem. Furthermore, supernova feedback in dwarf galaxies has the potential to `heat' the inner dark matter distribution, and transform a once cuspy halo into one with a central core.

Although the inclusion of baryonic physics in the context of CDM simulations has partly alleviated the claimed inconsistencies on small scales, perhaps the biggest challenge to the paradigm remains the lack of any robust detection of a candidate particle. Weakly interacting massive particles (WIMPs) could possibly be detected using both direct and indirect methods. Direct detection relies on nuclear recoil when WIMPs scatter off target nuclei \citep{Liu_2017}. These detection experiments have not yet had any success in finding a WIMP particle, but have put forward increasingly precise mass and cross-section constraints \citep{Aprile_2016, Tan_2016, Akerib_2017, Liu_2017}. Indirect detection of WIMPs requires observing the energy emission of dark matter annihilation. Energy emission anomalies, particularly in the Galactic centre, have been observed and could be attributed to annihilating dark matter \citep{Goodenough_2009,Hooper_2011,Calore_2015,Daylan_2016}, but recent claims suggest that the anomalous emission may not be related to dark matter at all, and may instead be sourced by an unresolved population of point sources in the Galactic centre \citep[e.g.][]{Abazajian_2011,Petrovic_2015,Lee_2016}. 

With no robust detection of a WIMP-like dark matter particle to date, there has been an increasing focus on well-motivated alternatives to the CDM model. One such example is Warm Dark Matter (WDM), in which dark matter particles were near relativistic in the early Universe, resulting in their free streaming that smoothed out density perturbations. This manifests as a cutoff in the initial power spectrum of matter fluctuations, and a suppression in the formation of low-mass structure; specifically, in the regime of dwarf galaxies \citep[e.g.][]{Maccio_2010,Kennedy_2014,Bose_2017,Bozek_2019}. Furthermore, the onset of structure formation in the early universe is also delayed with respect to a CDM universe. A leading candidate for WDM is the resonantly-produced sterile neutrino \citep{Asaka_2005,Abazajian_2006,Boyarsky_2009,Adhikari_2017}, and is the model we focus on in this work. Unlike WIMPs, the energy range for detecting these particles is in the X-ray, and is a result of particle decay as opposed to annihilation. Recently, there have been a number of claimed detections of X-ray emission that could possibly be attributed to the decay of sterile neutrinos with a rest mass of around 7 keV \citep{Boyarsky_2014,Bulbul_2014}, although alternative explanations are possible \citep[e.g.][]{Malyshev_2014,Anderson_2015,RiemerSorensen_2016,Dessert_2020}; this is the model of interest in the present work. Importantly, the particular variant of the 7 keV sterile neutrino that we consider is consistent with the lower limit on WDM-like particles derived from measurements of the Lyman-$\alpha$ forest \citep[][]{Viel_2013,Irsic_2017}, and has a small-scale cutoff similar to that of a thermal relic WDM particle with a rest mass of 3.3 keV. 

In the early universe ($z\gtrsim4$), the nature of the dark matter particle is most apparent since haloes in a WDM universe collapse later with respect to haloes in a CDM universe \citep[e.g.][]{Colin_2000,Bode_2001,Lovell_2012,Bose_2015}. The upcoming James Webb Space Telescope (JWST) will revolutionise the field of early galaxy formation and evolution. This new telescope will detect and characterise the first galaxies thanks to its (near-)infrared capabilities. Besides constraining the baryonic physics of galaxy formation at early cosmic times, it remains unclear whether JWST can also constrain the nature of dark matter by tracing the buildup of the first galaxies. Previous work has suggested that high-redshift observations by the JWST could provide strong constraints on the nature of dark matter \citep[e.g.][]{Schultz_2014,Dayal_2015,Maio_2015,Bose_2016}, particularly using the abundance of the faintest galaxies expected to be detected by the telescope at these early epochs.

However, we currently lack a complete picture of how baryonic and dark matter physics interplay with each other. For example, hydrodynamical simulations have shown that both CDM and sterile neutrino cosmologies produce equally plausible solutions to the small-scale puzzles in the Local Group \citep[e.g.][]{Lovell_2017}. Similarly, baryonic processes involved in the formation and evolution of galaxies at high redshifts ($z\gtrsim4$) are not well understood, possibly blurring the differences between CDM and WDM in this regime. In this work, we provide a quantitative analysis of how well CDM and WDM models may be differentiated by studying galaxies at high redshift with JWST-like observations. In particular, we explore the degeneracy between variations of baryonic physics and changes in the dark matter model. We consider two simulations -- one with CDM, and one with a 7 keV sterile neutrino -- to assess the impact of a primordial power spectrum with a small-scale cutoff. Throughout the remainder of this work, we refer to the sterile neutrino model simply as `WDM'. When varying our baryonic parameters, we test two different star formation efficiencies in our galaxy evolution model.

The layout of this paper is as follows. In Section~\ref{sec:methods}, we describe the galaxy evolution model (Section~\ref{sec:model}) and the simulations (Section~\ref{sec:dm_model}) used in this work, including the nature of the variations we explore in the latter (Section~\ref{sec:efficiency_calib}). Our main results are presented in Section~\ref{sec:results}. We then reflect on these results and discuss limitations of our model in Section~\ref{sec:discussion}. We summarise our work in Section~\ref{sec:conclusions}.

\section{Methods}
\label{sec:methods}

We adopt the galaxy evolution model introduced in \citet{Tacchella_2018} for exploring the degeneracy between the star formation efficiency and the underlying dark matter model. In the following section we summarise the dark matter simulation and the calibration of the star formation efficiency.

\subsection{Model Framework}
\label{sec:model} 

Following the \citet{Tacchella_2018} model, we assume that formation and evolution of galaxies follows a tight relationship with the build up of dark matter haloes. Specifically, we assume that the star formation rate (SFR) of each halo can be derived from the dark matter accretion rate and the star formation efficiency, $\varepsilon(M_{\rm halo})$, where $\varepsilon$ depends only on halo mass:

\begin{equation}
\begin{split}
    \mathrm{SFR}(M_{\rm halo},z) & = \varepsilon(M_{\rm halo}) \times \dot{M}_{\rm gas} \\
 & = \varepsilon(M_{\rm halo}) \times f_{\rm b} \times \widetilde{\frac{\mathrm{d}M_{\rm halo}}{\mathrm{d}t}}(M_{\rm halo},z),
\end{split}
\label{eq:SFR}
\end{equation}

\noindent
in which the rate of infalling baryonic mass ($\dot{M}_{\rm gas}$) is proportional to the mass accretion rate of the halo, rescaled by the universal baryon fraction $f_{\rm b}=\Omega_{\rm b}/\Omega_{\rm m}=0.167$. The term $\widetilde{\frac{\mathrm{d}M_{\rm halo}}{\mathrm{d}t}}$ is the delayed and smoothed accretion of dark matter onto its halo (see \citealt{Tacchella_2018} for further details). In this model, the only parameter that needs calibration is the star formation efficiency, $\varepsilon(\mathrm{M_{\rm halo}})$, which describes how efficiently gas is converted into stars, and encapsulates the complicated baryonic processes such as gas cooling, star formation, and various feedback processes into a single parameter. The function is constrained at $z=4$ by using observations of the UV luminosity function (UVLF) of galaxies. Calibrating the model at $z=4$ alone is able to reproduce observations of the UVLF at higher redshifts \citep{Tacchella_2013,Tacchella_2018}; we describe this procedure in more detail in Section~\ref{sec:efficiency_calib}. This model framework has also been applied to study the escape fraction distribution within the high-$z$ galaxy population, putting forward that bright galaxies may also contribute significantly to reionisation \citep{Naidu_2020}.

\begin{figure*}
	\includegraphics[width=\textwidth]{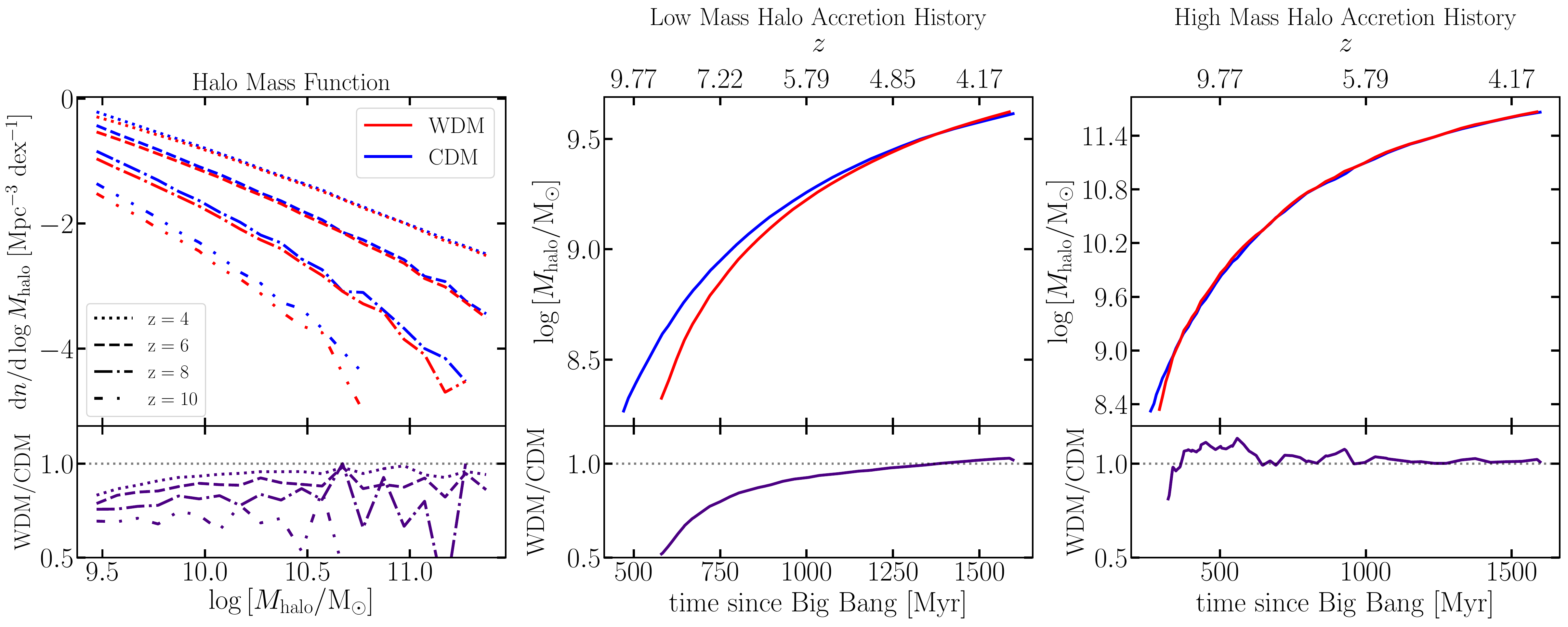}
    \caption{Halo mass function and dark matter accretion for the CDM and WDM models. The left panel shows the number of haloes at each halo mass bin, with WDM haloes shown in red and CDM haloes shown in blue. The centre and right panels show median accretion histories for low and high mass haloes. Low mass haloes are those with $10^{9}~ \mathrm{M_\odot} < M_\mathrm{{halo}} < 10^{10}~\mathrm{M_\odot}$ and high mass haloes are those with $M_{\mathrm{halo}} > 10^{11.5} ~\mathrm{M_\odot}$ at $z=4$. The expected trends in accretion histories are observed, with high mass haloes having similar accretion histories with only minor delays in WDM structure formation and low mass haloes having noticeable differences in accretion histories between WDM and CDM with a significant delay and a higher accretion rate observed in WDM that will eventually catch up to CDM at $z=4$. The lower subpanels show the ratio of WDM to CDM for the corresponding quantity.}
    \label{fig:dm}
\end{figure*}

\subsection{Dark Matter}
\label{sec:dm_model}

The dark matter haloes are obtained from the \textsc{Color} $N$-body simulation \citep{Hellwing_2016, Sawala_2016} which has a computational volume of (100 Mpc)$^{3}$. The simulation follows the evolution of $1620^3$ particles with an effective particle mass of $8.8 \times 10^6$ M$_\odot$.

Halo merger histories constructed from the \textsc{Color} simulations act as the foundation upon which our galaxy evolution model is built. The simulations adopt cosmological parameters from the 7-year Wilkinson Microwave Anistropy Microwave Probe (WMAP7, \citealt{Komatsu_2011}): $\Omega_{m} = 0.272, \Omega_\Lambda = 0.728, h = 0.704, n_s= 0.967,$ and $\sigma_8 = 0.81$. We model the impact of WDM (i.e., a 7 keV sterile neutrino) by modifying the initial conditions of the simulation using an initial power spectrum with a small-scale cutoff; for more details, we refer the reader to \cite{Bose_2015}.

In Fig.~\ref{fig:dm}, we compare statistics of the halo population extracted from our CDM and WDM simulations. Specifically, we plot the halo mass function at $z=4, 6, 8$ (left panel), and $10$, as well as mass accretion histories for high mass ($M_{{\rm halo}} > 10^{11.5}~{\rm M}_\odot$, centre panel) and low mass ($10^9~{\rm M}_\odot<M_{{\rm halo}}<10^{10}~{\rm M}_\odot$, right panel) haloes. In each case, we can see that the largest differences between WDM and CDM occur in low mass haloes while high mass haloes are nearly indistinguishable in the two models. This is unsurprising since the characteristic cutoff mass scale for the WDM model considered in this work only suppresses the formation of haloes less massive than $\sim10^{10}~{\rm M}_\odot$. 

Fig.~\ref{fig:dm} also shows that the differences between the two models are more pronounced at earlier times, both in terms of the abundance of haloes, as well as in the mass assembly of dwarf galaxy haloes. Given the mass resolution of our simulation, the earliest progenitors of WDM haloes begin to collapse roughly 100 Myr after their counterparts in CDM. The early phase of the mass assembly is also more rapid in the WDM scenario, which is evidenced by the steeper gradient of the red curve in the centre panel. In Section~\ref{sec:results}, we will investigate the extent to which these differences propagate to properties of the galaxy population that may be measurable by JWST.

\subsection{Star Formation Efficiency Calibrations}
\label{sec:efficiency_calib}

\begin{figure}
	\includegraphics[width=\columnwidth]{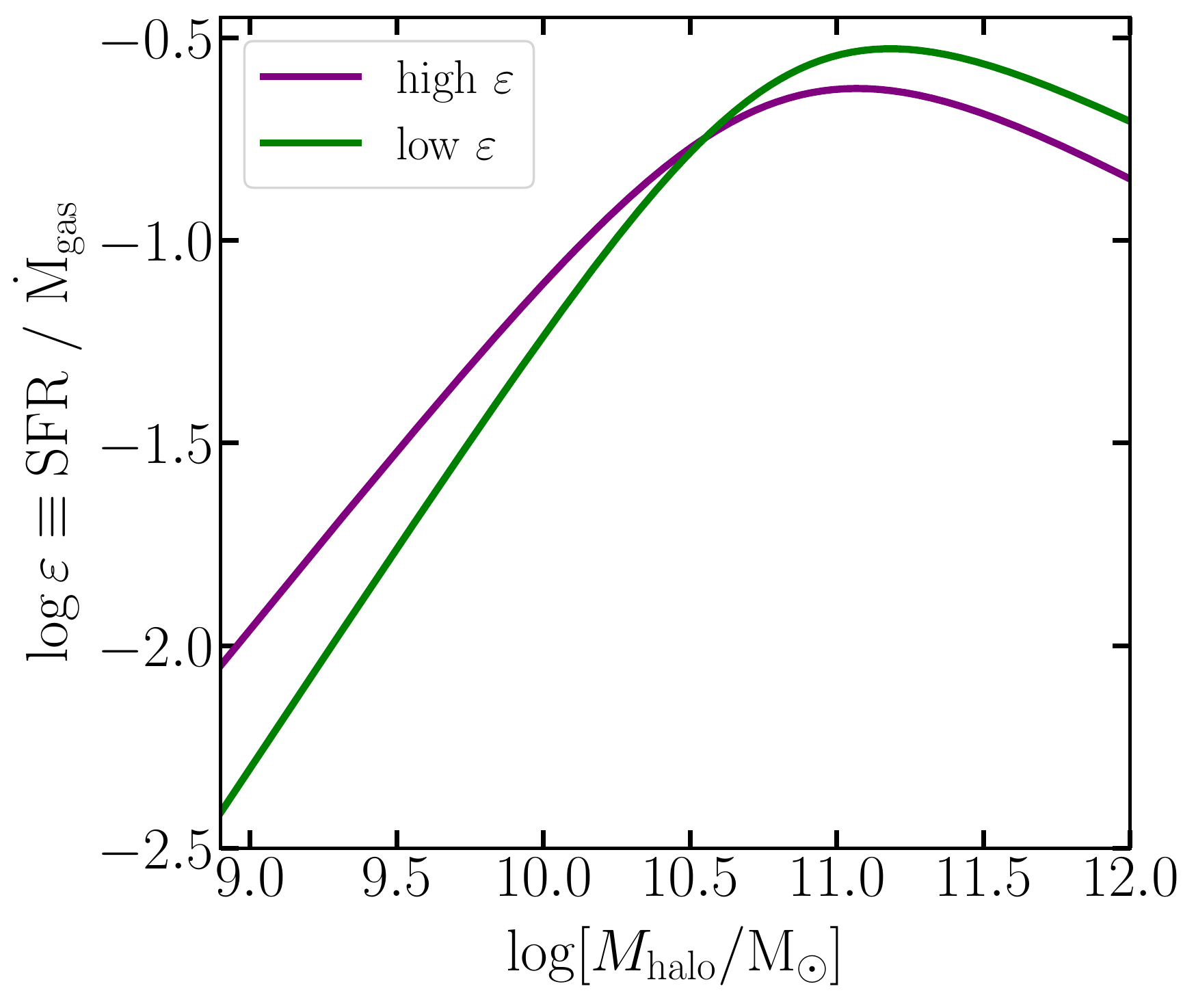}
    \caption{High and low star formation efficiency ($\varepsilon$) models. $\varepsilon$ describes how efficiently gas is turned into stars (see Eqn. \ref{eq:epsilon}) and is assumed to depend solely on halo mass $M_{\rm halo}$. The high and low efficiency models are motivated by adopting the \citet{Meurer_1999} and SMC dust attenuation prescription \citep{Gordon_2003}, respectively. We use these two different star formation efficiency models to explore the degeneracy between baryonic physics and the dark matter model (WDM versus CDM).}
    \label{fig:epsilon}
\end{figure}

\begin{figure*}
	\includegraphics[width=\textwidth]{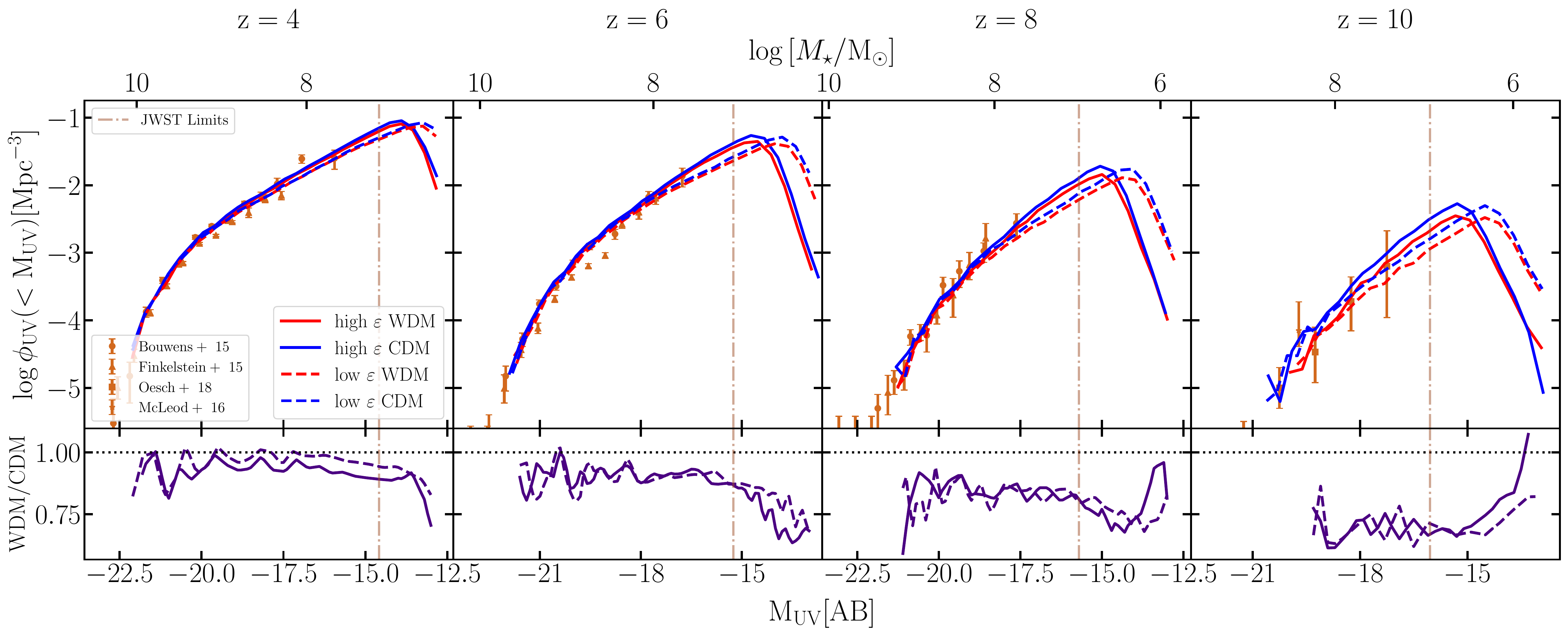}
    \caption{UV luminosity function (UVLF) at $z=4,6,8, \mathrm{and\ } 10$. The red lines represent the WDM model and the blue lines represent CDM. The dashed and solid lines are those with lower and higher star formation efficiency, respectively. The bottom panel shows the ratios between WDM and CDM. Observational data shown in orange are taken from \citet{J_Bouwens_2016}; \citet{Finkelstein_2015}; \citet{Oesch_2018}; and \citet{McLeod_2016}. The top x-axis corresponds to the stellar masses at those UV magnitudes. The vertical lines are the JWST sensitivity limits for an HUDF-like survey \citep{Yung_2018}. Only at $z=10$ do the WDM and CDM lines begin to diverge and show visual differences in UVLF. These differences are small, especially above the observational limit of JWST. Differences between the two efficiencies, however, can be seen in faint objects at $z=6$.}
    \label{fig:UVLF}
\end{figure*}

We calibrate the star formation efficiency via the UVLF at $z=4$. In order to predict UV luminosities from our model, the spectral energy distributions (SEDs) are calculated by producing star formation histories (SFHs) from the SFRs derived in the model framework. While we perform the calibration using the CDM model, we find that the same calibration is also valid for the WDM model since the halo mass function and accretion histories are nearly indistinguishable at $z=4$ (Fig.~\ref{fig:dm}). We confirm this also in Section \ref{sec:uvlf}. The SED for each halo is determined from the Flexible Stellar Population Synthesis code \citep[FSPS\footnote{\href{https://github.com/cconroy20/fsps}{https://github.com/cconroy20/fsps}};][]{Conroy_2009}. MIST isochrones \citep{2016ApJ...823..102C} and the MILES stellar library \citep{2015MNRAS.449.1177V} are used to determine stellar properties. In  deriving the SEDs, we assume a metallicity of $0.02\ \mathrm{Z_\odot}$ for all galaxies and an IMF by \citet{1955ApJ...121..161S}.

In order to calibrate $\varepsilon(\mathrm{M_{\rm halo}})$ to observations, dust attenuation has to be taken into account in the model. We adopt the same attenuation model as in \citet[][see also \citealt{Tacchella_2013, Mason_2015, Vogelsberger_2020}]{Tacchella_2018} by adopting the observed relation between the UV luminosity and the UV continuum slope from \citet{Bouwens_2014}. We then adopt two different dust attenuation laws in order to derive two different star formation efficiencies. Specifically, we use the \citet{Meurer_1999} and the Small Magellanic Cloud (SMC) relation from \citet{Gordon_2003}.

The use of two different dust attenuation models leads to different $\varepsilon(\mathrm{M_{h}})$ functions that match the observed UVLF at $z=4$ equally well (see Fig. 2 in \citealt{Tacchella_2018} and Fig.~\ref{fig:UVLF} in this paper). The functions are parameterised with a relation from \citet{Moster_2010}:

\begin{equation}
    \varepsilon(M_{h}) = 2\varepsilon_0\left( \left[\dfrac{M_{{\rm halo}}}{M_c}\right]^{-\beta} + \left[\dfrac{M_{{\rm halo}}}{M_c}\right]^{\gamma}
    \right),
\label{eq:epsilon}
\end{equation}

where $\epsilon_0$, $M_c$, $\beta$, and $\gamma$ are free parameters to be calibrated. For the low $\varepsilon$ model \citep[denoted `Z-const' in][]{Tacchella_2018}, the \citet{Meurer_1999} dust attenuation relation is used. For this model, $(\varepsilon_0, M_c, \beta, \gamma) = (0.37, 7.10\times10^{10}, 1.09, 0.36)$. The high $\varepsilon$ model \citep[denoted `SMC' in][]{Tacchella_2018} uses the SMC IRX-$\beta$ relation \citep{Gordon_2003} and has $(\varepsilon_0, M_c, \beta, \gamma) = (0.22, 6.30\times10^{10}, 0.89, 0.40)$. The distinction between these different efficiencies is most notable at lower halo masses as seen in Fig.~\ref{fig:epsilon}. In what follows, we refer to each of these scenarios as the high and low $\varepsilon$ models based on their behaviour at lower halo masses, where dark matter haloes are most abundant (Fig.~\ref{fig:dm}).

\section{Results}
\label{sec:results}

In the following subsections, we present our results on how the baryonic prescription (by varying the star formation efficiency $\varepsilon(M_{\rm halo})$) is degenerate with the underlying dark matter physics (CDM versus WDM) by showing differences in key observational probes of early galaxy formation, including the UVLF, cosmic SFR and stellar mass density, and stellar mass histories. We focus on the redshift range $z=4-12$, which is of great interest for the upcoming JWST.

\subsection{UV luminosity function (UVLF)}
\label{sec:uvlf}

Fig.~\ref{fig:UVLF} shows the resulting UVLF from our model at $z=4$ as well as higher redshifts. As described in Section~\ref{sec:efficiency_calib}, the different star formation efficiencies are a result of two different dust attenuation prescriptions. For the low $\varepsilon$, we use the SMC prescription \citep{Gordon_2003} while the high $\varepsilon$ uses the prescription presented in \citet{Meurer_1999}. Both star formation efficiency models are calibrated to observations at $z=4$, which leads to the good agreement between model and observations in Fig.~\ref{fig:UVLF}. The high $\varepsilon$ model (solid lines), which assumes a higher star formation efficiency in low-mass haloes (see Fig.~\ref{fig:epsilon}), predicts a steeper faint-end slope of the UVLF than the low $\varepsilon$ model (dashed lines). This is expected since the higher star formation efficiency in low mass haloes leads to a higher SFR and larger count of objects at low luminosity. While the efficiencies are calibrated at $z=4$, the same $\varepsilon(M_{\rm halo})$ still shows good agreement with observations at higher redshifts. This indicates that the buildup of dark matter haloes together with a simple dust prescription can largely explain the evolution of the UVLF with cosmic time, and there is no need to invoke a redshift-dependent star formation efficiency \citep[see also][]{Behroozi_2012,Tacchella_2013}.

Focusing now on the effect of different dark matter models on the UVLF, we plot in Fig.~\ref{fig:UVLF} the CDM and WDM predictions in blue and red, respectively. At $z=4$, we see that all models agree well with one another, and with observations. This is to be expected as the models were calibrated to the data at $z=4$. At higher redshifts, all four models continue to show good agreement with observations. However, differences are noticeable in fainter objects and at higher redshifts ($z=8,10$). Similar to trends in Fig.~\ref{fig:dm}, we expect that differences between WDM and CDM would be most prominent at higher redshifts and for lower mass haloes. Changes between star formation efficiency and dark matter model, however, have similar effects, which makes the UVLF of high $\varepsilon$ WDM and low $\varepsilon$ CDM difficult to distinguish.

These trends can be seen in the UVLF plots. However, these largest differences will likely not be able to be observed as indicated by the vertical dashed lines in the figure, which show the JWST observational limit assuming an HUDF-like survey \citep{Yung_2018, 2018ApJS..236...33W}. For objects brighter than the JWST sensitivity limit, the four models show little deviation at $z=4$. At $z=6$ and $z=8$, there are more prominent differences between efficiency models, but differences between dark matter models remain small. At $z=10$, we begin to see larger differences between WDM and CDM, but differences between the high $\varepsilon$ WDM and low $\varepsilon$ CDM continue to be small.

\subsection{Cosmic Star Formation Rate Density and Stellar Mass Density}

\begin{table}
	\centering
	\caption{Fits for SMD and SFRD shown in Fig.~\ref{fig:SMD} with $\log({\mathrm{SMD}})=m\times z +\delta$ where $z$ is redshift. We also include fits to the SMD obtained by \citet{Dayal_2015}.}
	\label{tab:fits}
        \begin{tabular}{|p{2.1cm}|p{.85cm}|p{.8cm}|p{.85cm}|p{.8cm}|}
         \hline
         & \multicolumn{2}{|c|}{SMD Fits} & \multicolumn{2}{|c|}{SFRD Fits}
         \\
         \hline
          & $m$ & $\delta$ & $m$ & $\delta$\\
         \hline
         high $\varepsilon$ CDM &  -0.45& 9.55&-0.47&1.03\\
         high $\varepsilon$ WDM &  -0.48& 9.67& -0.52&1.28\\
         low $\varepsilon$ CDM &  -0.50& 9.71& -0.52& 1.27\\
         low $\varepsilon$ WDM &  -0.53& 9.86& -0.59& 1.66\\
         \hline
         \citet{Dayal_2015} CDM &  -0.44& 9.86& -------&-----\\
         \citet{Dayal_2015} 1.5keV WDM &  -0.63& 11.27& -------&-----\\
         \hline
        \end{tabular}
\end{table}

\begin{figure}
	\includegraphics[width=\columnwidth]{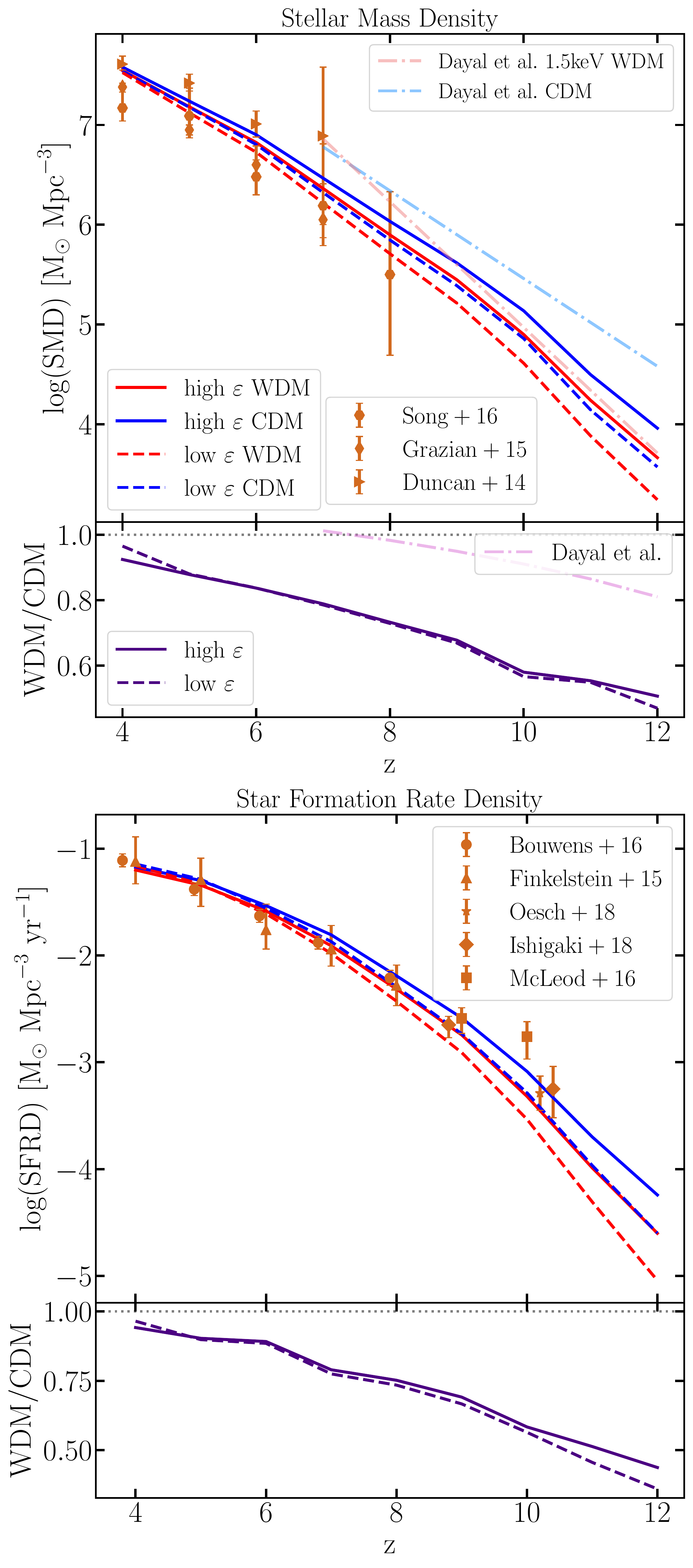}
    \caption{\textit{Top:} Cosmic stellar mass density (SMD) at $z = 4-12$. Observational data is taken from \citet{Duncan_2014}; \citet{Grazian_2015}; and \citet{Song_2016}. Figure also includes relation between stellar mass density and redshift cited in \citet{Dayal_2015}, although we integrate down to a stellar mass of $10^7\ \mathrm{M_\odot}$ while \citet{Dayal_2015} integrates down to a magnitude $-16.5\ \mathrm{M_{UV}}$. Summary of linear fits and comparisons to \citet{Dayal_2015} can be found in Table \ref{tab:fits}. \textit{Bottom:} Cosmic star formation rate density (SFRD) at $z = 4-12$. Observational data is from \citet{J_Bouwens_2016}; \citet{Finkelstein_2015}; \citet{Oesch_2018}; \citet{Ishigaki_2015}; and \citet{McLeod_2016}.  Bottom panels show ratios between WDM and CDM. It is evident that looking at both cSMD and cSFRD, the distinction between high $\varepsilon$ WDM and low $\varepsilon$ CDM is small and both lines follow similar trends at all redshifts.}
    \label{fig:SMD}
\end{figure}

\begin{figure*}
	\includegraphics[width=\textwidth]{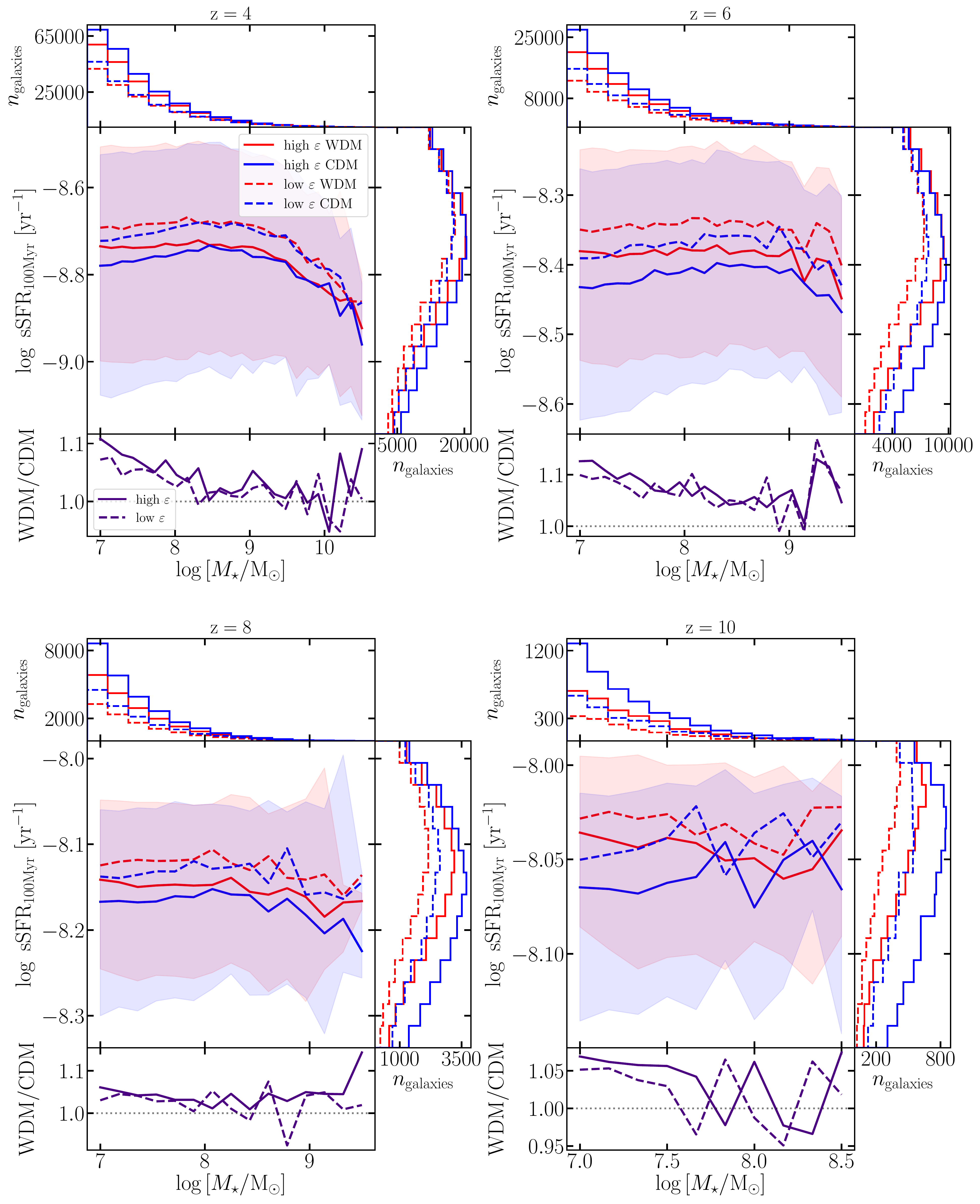}
    \caption{Specific star formation rates ($\mathrm{sSFR}=\mathrm{SFR}/M_{\star}$) as a function of mass at $z=4,6,8, \rm{and}\ 10$. The SFR is averaged over 100 Myr in all diagrams. Histograms along the $x$ and $y$-axis show density of haloes at each stellar mass and sSFR. Bottom panels show ratios between WDM and CDM for each efficiency model. For clarity, we only show scatter for high $\varepsilon$ models, however, scatter for low $\varepsilon$ models follows a similar trend. At $z=4$, the distinction between the high and low $\varepsilon$ models is evident, while the differences between WDM and CDM are small. As redshift increases, the distinction between high $\varepsilon$ WDM and low $\varepsilon$ CDM declines. At  $z=10$, all four models are largely indistinguishable, although the median ratio between WDM and CDM of any given efficiency is larger than the ratio between high $\varepsilon$ WDM and low $\varepsilon$ CDM.}
    \label{fig:sSFR}
\end{figure*}

\begin{figure*}
	\includegraphics[width=\textwidth]{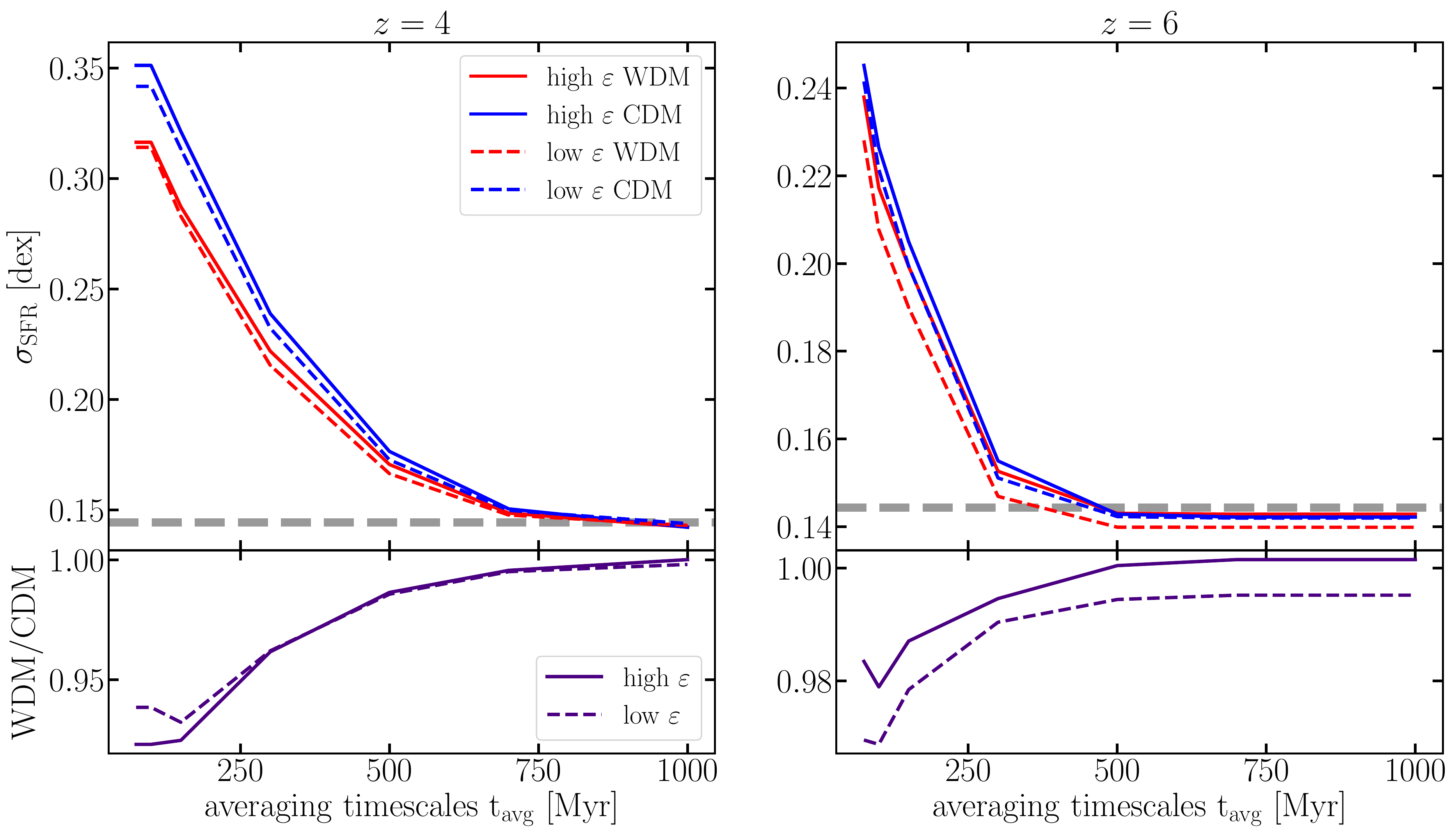}
    \caption{Scatter of the star formation rate (SFR) at fixed $M_{\star}$ (i.e. of the main sequence) $\sigma_{\rm SFR}$ as a function averaging timescale of the SFR. Scatter is largely independent of efficiency and depends more so on the underlying dark matter model. Specifically, in our model framework, dark matter accretion rates and therefore star formation rates are smoother (i.e. less bursty) in WDM with respect to CDM, leading to less scatter in SFR. }
    \label{fig:scatter}
\end{figure*}

Another important observational constraint on the formation of galaxies in the early Universe is the measurement of how quickly stellar mass in galaxies is formed. In Fig.~\ref{fig:SMD} we show the evolution of cosmic stellar mass density (SMD) and cosmic star formation rate density (SFRD) from $z=4$ to $z=12$. For both plots, most observations match all four models for $z\lesssim 8$. Most notably, we see that high $\varepsilon$ WDM and low $\varepsilon$ CDM follow surprisingly similar trends for both SMD and SFRD. In Table~\ref{tab:fits}, we show the gradient and intercept of linear fits to all four models from $z=4$ to $z=12$, as well as comparisons to values presented in \citet{Dayal_2015}, who investigated similar quantities in CDM and a 1.5 keV thermal relic WDM model\footnote{We remind the reader that the 7 keV sterile neutrino we study in this paper maps to a 3.3 keV thermal relic WDM particle.}. The high $\varepsilon$ WDM and low $\varepsilon$ CDM fits have slopes that are most similar, and the slope of our high $\varepsilon$ CDM model agrees well with the slope for CDM predicted by \citet{Dayal_2015}.

Importantly, the similarity between the high $\varepsilon$ WDM and low $\varepsilon$ CDM indicates that the differences between WDM and CDM models can be replicated by altering baryonic physics. Specifically, increasing the star formation efficiency in low-mass haloes in a WDM universe, where halo accretion rates (and therefore also gas accretion rates) in the early universe are lower than in CDM (Fig.~\ref{fig:dm}) mimics CDM with a lower star formation efficiency. Therefore, it is difficult to constrain the dark matter model without independently constraining the baryonic physics, i.e. the star formation efficiency in our model.

\subsection{The distribution of star formation rates}

In this section we compare WDM and CDM models of different efficiencies by studying specific star formation rates ($\mathrm{sSFR}=\mathrm{SFR}/M_{\star}$) as a function of stellar mass and redshift. This relation between SFR and $M_{\star}$ is sometimes called the star-forming main sequence -- a key galaxy scaling relation \citep[e.g.][]{2007ApJ...660L..43N, 2014ApJ...795..104W}.

\begin{figure*}
	\includegraphics[width=\textwidth]{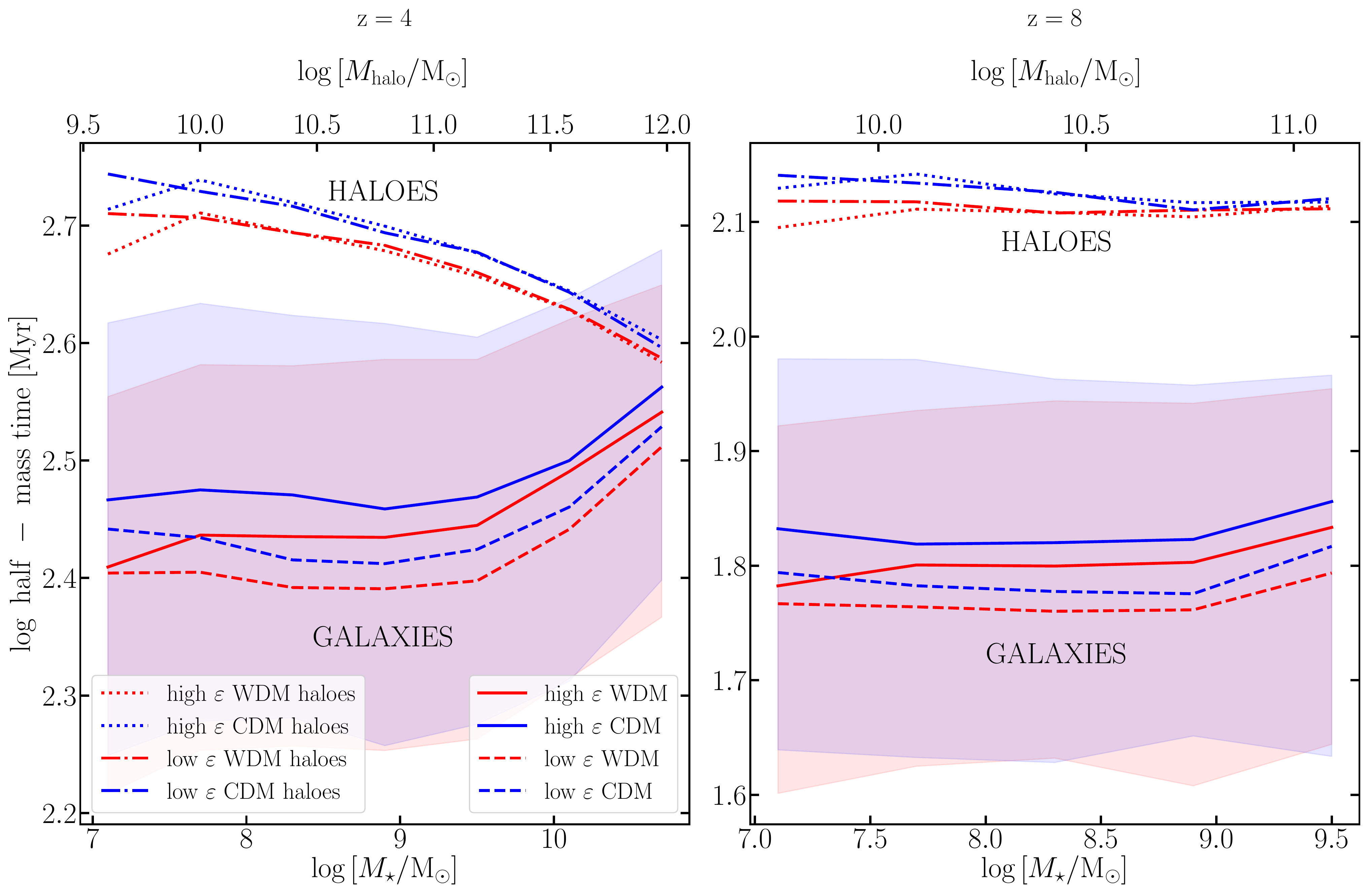}
    \caption{Ages of haloes and galaxies at $z=4$ and $z=6$. Bottom axes show stellar masses while top axes show corresponding halo masses. For clarity, we only show scatter for high $\varepsilon$ models, however, scatter for low $\varepsilon$ models follows a similar trend. At $z=4$, halo ages start to decrease while galaxy ages increase in haloes with a mass greater than $10^{11}\ \mathrm{M_\odot}$. The trend is a result of both $\varepsilon(\mathrm{M_h)}$ functions reaching a peak and decreasing for haloes with masses $10^{11}\ \mathrm{M_\odot}$ or greater (see Fig.~\ref{fig:epsilon}). These younger haloes grow massive faster, but star formation is no longer as efficient, leading to galaxies with older stellar populations. No such trend is observed at $z=8$ because these haloes are not yet massive enough to have a decreasing star forming efficiency.}
    \label{fig:ages}
\end{figure*}

In Fig.~\ref{fig:sSFR}, we show sSFR as a function of stellar mass for haloes at $z = 4,6,8,\rm{and}\ 10$, with the density of haloes shown using rotated histograms along the axes. Here, the SFR is measured over a time interval of 100 Myr. At all redshifts, the sSFR is roughly constant as a function of $M_{\star}$, consistent with a linear relation between SFR and $M_{\star}$ (see also \citealt{Tacchella_2018}). This means that low and high mass galaxies double their mass at the same rate. At $z=4$, galaxies typically double their mass every $\sim600$ Myr, while galaxies at $z=10$ roughly double their mass every 100 Myr. Taking a closer look at the high-mass end at $z=4$, we can see that the sSFRs decrease for galaxies with $M_{\star}>10^9\ \rm{M}_\odot$ for all four models due to the decreasing star formation efficiency in high mass haloes seen in Fig.~\ref{fig:epsilon}. This effect is sometimes referred to as the `bending' of the star-forming main sequence. We examine this further in Section~\ref{sec:ages}.

Comparing dark matter models in Fig.~\ref{fig:sSFR}, we see that for haloes at lower redshifts, the distinction between WDM and CDM models is small, and the median difference between the two, taken over all galaxy masses, is consistently under 5\%. Comparing efficiency models, however, shows a significant difference between high and low $\varepsilon$, with a median ratio over all masses that is more than 12\%. At higher redshift, the median ratios between dark matter models stay roughly the same, never rising above 7\%. The median ratios between the two efficiency models, however, decrease steadily to less than 6\% at $z=10$. Most notably, at  $z=6$, the median ratio between high $\varepsilon$ WDM and low $\varepsilon$ CDM drops from 9\% at $z=4$ to 3\%. The ratio remains small at high redshifts as well. No other combination of models produces populations with a similar decrease in ratios.

In trying to further determine how we can break the degeneracy between the star formation efficiency and dark matter model, we investigate the scatter in SFR ($\sigma_{\mathrm{SFR}}$) at fixed $M_{\star}$ over increasing SFR averaging timescales. This scatter of the star-forming main sequence encodes information regarding the regulation of star formation within galaxies \citep{2016ApJ...832....7A, 2016MNRAS.457.2790T, 2019MNRAS.484..915M, 2019MNRAS.487.3845C}. In particular, as shown in \citet{2019MNRAS.487.3845C}, averaging the SFR over different timescales is a measure of the temporal power spectrum of SFR fluctuations, which in turn can be linked to physical processes within galaxies \citep{2020MNRAS.497..698T,2020MNRAS.498..430I}.

Fig.~\ref{fig:scatter} shows the scatter in this relation at $z= 4$ and $6$, where the SFR is averaged over time intervals ranging from 75 to 1000 Myr. The lower limit is given by the snapshot spacing of the numerical simulation we have used. We find that the scatter $\sigma_{\mathrm{SFR}}$ decreases towards higher stellar masses and higher redshift. Most importantly, the figures show that scatter is largely independent of efficiency, but is sensitive to the dark matter model. The difference between the WDM and CDM amounts to $\sim10\%$; specifically, the scatter is smaller in the WDM case. This can be directly understood by considering that the dominant mode of mass growth in WDM haloes is via the smooth accretion of matter, leading to smoother, less bursty star formation. This shows that the observational plane of the star-forming main sequence could potentially distinguish between WDM and CDM within our model framework. 

Averaging the SFR over longer timescales leads to a decrease in $\sigma_{\mathrm{SFR}}$ as expected for a star formation history that decorrelates over a certain timescale \citep{2019MNRAS.487.3845C}. The faster decrease of $\sigma_{\mathrm{SFR}}$ at higher redshifts can be understood by the bursty nature star formation during these epochs, which decorrelates the star formation history on shorter timescales. The scatter does not decrease to 0 because we are probing a finite mass bin with a width of 0.5 dex. In the limit where the averaging timescale is the age of the universe, $\sigma_{\mathrm{SFR}}$ is given by the variance of a uniform distribution with a width of 0.5 dex, i.e. $\sqrt{1/12\cdot(0.5)^2}=0.144$. This limit is indicated with a horizontal dashed grey line in Fig.~\ref{fig:scatter}. Indeed, we find that $\sigma_{{\rm SFR}}$ largely asymptotes to this value in the limit of large averaging timescales.

\subsection{Stellar and Halo Ages}
\label{sec:ages}

After analysing the recent SFR in WDM and CDM galaxies, we now study the ages of high-$z$ galaxies and their parent haloes. In particular, we define the age, $t_{50}$, to be the look-back time when 50\% of the mass was formed. In studying halo and galaxy ages, we find that changing the star formation efficiency results in an interesting relationship between halo and stellar ages in high mass haloes. 

Fig.~\ref{fig:ages} shows galaxy ages and their scatter, as well as halo ages as functions of total stellar mass. We show the corresponding halo masses along the top axes. The stellar-to-halo mass relation is similar in all four models. At $z=4$, high mass haloes tend to decrease in age, which is well-known since the most massive haloes have assembled most recently. On the other hand, the stellar populations of these massive haloes tend to be older. This trend is a result of the efficiency function shown in Fig.~\ref{fig:epsilon}. In both the high and low $\varepsilon$ models, the functions reach a peak and begin decreasing at halo masses of $10^{10.5}-10^{11}\ \rm{M}_\odot$. In Fig.~\ref{fig:ages}, it is at these halo masses where younger haloes have less efficient star formation, resulting in older stellar populations. These younger haloes grew to be massive faster, but at some point their masses resulted in a less efficient conversion of gas into stars. This trend is not seen at $z=8$, since there are no haloes massive enough to have a decreasing star formation efficiency. Overall, this picture is consistent with the trend of constant sSFR with $M_{\star}$ (see Fig.~\ref{fig:sSFR}), which indicates that all galaxies double their mass over roughly the same timescale (only the most massive galaxies at $z=4$ have a lower sSFR with respect to their lower-mass counterparts).

Comparing the low and high $\varepsilon$ models, we find that a lower efficiency leads to younger stellar populations. This can again be explained by looking at Fig.~\ref{fig:epsilon}, where the slope at low halo masses is steeper for the lower efficiency model. In order to achieve the same stellar mass at a certain redshift in both the low and high $\varepsilon$ models, the galaxies in the former case need to catch up by forming more stars in recent times.

A nearly identical trend can be found when studying $t_{10}$, which is the lookback time when 10\% of the total stellar mass was formed. In this case, it takes stars and haloes less time to accumulate 10\% of their mass, so ages are shifted upwards, but no other systematic differences in the trends are present.

\section{Discussion}
\label{sec:discussion}

\subsection{The difficulty of constraining the dark matter model}

We have adopted an empirical model of galaxy evolution to explore the degeneracy between baryonic processes (parameterised by the star formation efficiency) and the underlying dark matter model (CDM versus WDM). We use two different efficiencies (low and high $\varepsilon$), which were derived from calibrating the UVLF using two different dust attenuation models. 

Interestingly, we find that there are observational characteristics of galaxies that are unaffected by altering the star formation efficiencies, and mainly probe the dark matter model. Specifically, the scatter in star formation rates at fixed stellar mass (i.e. the scatter of the star-forming main sequence) shows a dependence on the dark matter model that is stronger than the dependence on the efficiency of star formation (Fig.~\ref{fig:scatter}). A caveat of this statement is our galaxy evolution model assumes that the star formation rates are directly related to the dark matter accretion rate, which in the real universe might not be the case (see also Section~\ref{sec:limitations}). 

Previous work has studied the differences between WDM and CDM, in order to find observable differences between the two models. In \citet{Dayal_2015}, the authors examine the differences between CDM and three different WDM models ($m_x = 1.5, 3$ and 5 keV) at high redshifts ($z\gtrsim7$) using a semi-analytic model of galaxy formation. Similar to the model we implement, the authors calibrate their model to the UVLF and implement a redshfit independent star formation efficiency, although a single star formation efficiency, throughout their work. \citet{Dayal_2015} find that the 3 keV WDM (comparable to ours) and CDM models exhibit the smallest differences, but that notable differences exist at high redshifts and that these differences can be possibly be observed. Similar findings are presented by \citet{Maio_2015}, also between 3 keV and CDM and by \citet{Schultz_2014} between 2.6 keV WDM and CDM. We also find that observable differences exist between WDM and CDM when considering a single efficiency model. In fact, the SFRD trends shown by \citet{Maio_2015} agree well with the trends presented in Fig.~\ref{fig:SMD} between models of a given efficiency.

While differences between dark matter models of the same efficiency exist, we find that uncertainties in the treatment of baryonic physics -- namely, the efficiency of star formation in high redshift haloes -- may further cloud the differences between individual dark matter models. We have found that by varying a single poorly constrained parameter in our model, $\varepsilon\left(M_{{\rm halo}}\right)$, it is possible to completely eradicate the residual differences between WDM and CDM, which were not large to begin with. In particular, differences between WDM and CDM are minimised when considering two different star formation efficiencies. 

Our findings indicate that while the JWST may provide better constraints on astrophysical parameters and dark matter particle masses, its constraints on dark matter are unlikely to be conclusive. This is in contrast to predictions made in previous work \citep{Schultz_2014, Maio_2015, Dayal_2015}. 
Although we find that the JWST may not provide conclusive constraints on the particle nature of dark matter, we find possible means of better constraining star formation efficiencies of dark matter haloes during these early epochs. We find that the largest differences in efficiencies present themselves at lower redshift (Fig.~\ref{fig:sSFR}), and show interesting trends between galaxy and halo ages in high mass haloes (Fig.~\ref{fig:ages}). The JWST remains a powerful probe of galaxies in the high redshift Universe and will still provide valuable constraints on astrophysical parameters and baryonic processes.

\subsection{Limitations of the present work}
\label{sec:limitations}

The main limitation of our work is related to the simplicity of the \citet{Tacchella_2018} galaxy evolution model, which directly couples the  baryonic processes within galaxies to the dark matter haloes themselves. Specifically, we assume that the star formation rate in galaxies is related to the the dark matter halo accretion rate (Eq. \ref{eq:SFR}). As outlined in \citet{Tacchella_2018}, this assumption can be motivated from theoretical grounds, but it also works surprisingly well since this simple ansatz is able to accurately reproduce a large range of observations (including the redshift evolution of the UVLF and cosmic star formation rate density). 

Nevertheless, because of this direct coupling between the star formation rate and the dark matter accretion rate, our finding that the scatter of star formation rates at fixed stellar mass (i.e. the scatter of the star-forming main sequence) depends primarily on the dark matter model must be interpreted with care (Fig.~\ref{fig:scatter}). In our model framework, the WDM model is characterised by a smoother accretion rate relative to CDM and, hence, the star formation rate is less variable (less bursty). This results in a smaller scatter and a different dependence on the averaging timescale. However, the short-timescale star formation variability is probably influenced by a host of baryonic processes internal to a galaxy (e.g., \citealt{2014MNRAS.445..581H}, \citealt{2020MNRAS.497..698T}). In particular, \citet{2020MNRAS.498..430I} use the IllustrisTNG cosmological, hydrodynamical simulations \citep{Pillepich_2018} to show that the variability of the star formation rate is coherent with the dark matter accretion rate only on the longest ($\sim$ Gyr) timescales. Therefore, it would be of interest to verify whether our findings can be replicated in numerical models equipped with more complex star formation prescriptions.

There are additional limitations arising from the fact that we use an $N$-body simulation that has a resolution limit of $\sim2\times10^7 \rm{M_\odot}$ in halo mass and a modest computational volume of (100 Mpc)$^3$. However, these caveats have no significant impact on our main results. A potentially fruitful way forward may be offered by galaxy clustering: galaxies with certain properties (e.g. colour, star formation rate etc.) may be clustered differently in WDM and CDM. In principle, our model framework is well-equipped for a study of this kind since galaxies are ``painted'' onto an $N$-body simulation. However, this is beyond the scope of the present analysis and we postpone a detailed examination of this to a future investigation.

\section{Conclusions}
\label{sec:conclusions}
The upcoming JWST will revolutionise the field of early galaxy evolution. Beside advancing our understanding of baryonic processes related to galaxies, a big open question is whether we can also constrain the nature of dark matter itself using observations of the early phases of galaxy formation. In this work, we present our findings on the degeneracy between baryonic processes in galaxy evolution and the underlying dark matter physics. We focus on two different dark matter models (CDM and a 7 keV sterile neutrino, or WDM), both of which are viable dark matter candidates. We then use a simple empirical galaxy evolution model, which links the star formation rate in galaxies to the accretion rate of their parent dark matter haloes \citep{Tacchella_2018}. We find that by using two different star formation efficiencies (see Sec. \ref{sec:efficiency_calib} and Fig.~\ref{fig:epsilon}), we are able to reduce the differences between WDM and CDM, including the most prominent differences at high redshifts (see Figs. \ref{fig:SMD} and \ref{fig:sSFR}). Our main findings are summarised as follows:

\begin{enumerate}
    \item The most pronounced differences between WDM and CDM haloes are present at higher redshifts in low mass haloes (Fig.~\ref{fig:dm}).
    
    \item The UV luminosity function shows little deviation between the 4 different models (two star formation efficiencies per dark matter candidate) at the observational limits of JWST at higher redshifts. While there is a more prominent distinction between CDM and WDM at $z=10$ for a {\it given} efficiency, varying the star formation efficiency is able to eradicate these differences entirely. In particular, the high $\varepsilon$ WDM and low $\varepsilon$ CDM resemble each other quite closely (Fig.~\ref{fig:UVLF}).
    
    \item All four models (high $\varepsilon$ CDM, low $\varepsilon$ CDM, high $\varepsilon$ WDM, and low $\varepsilon$ WDM) match a wide range of measurements of the cosmic stellar mass and star formation densities. Most notably, low $\varepsilon$ CDM and high $\varepsilon$ WDM follow nearly identical trends for all redshifts ($z=4-12$). This is evident in Fig.~\ref{fig:SMD} as well as in the linear fits we present in Table \ref{tab:fits}.
    
    \item At $z=4$, we see large differences in the specific star formation rates for high and low $\varepsilon$ models of a given dark matter model, and only a small difference between dark matter models of the same efficiency. At higher redshift ($z\geq6$), as the differences between dark matter models increase, the differences between high $\varepsilon$ WDM and low $\varepsilon$ CDM are significantly smaller than those between dark matter models of the same efficiency (Fig.~\ref{fig:sSFR}).
    
    \item On the other hand, we find that the scatter in star formation rates at fixed stellar mass is independent of star formation efficiency and differences are attributed only to different dark matter models (Fig.~\ref{fig:scatter}).
    
    \item Finally, we show that the star formation efficiency models shown in Fig.~\ref{fig:epsilon} lead to a decreasing star formation efficiency in high mass haloes. As a result, more massive, younger haloes tend to have older stellar populations. This trend can only be seen at $z=4$, as haloes massive enough cause this effect are not found in our simulations at higher redshift. (Fig.~\ref{fig:ages}).

\end{enumerate}

While our results suggest that there are indeed strong degeneracies between implementations of baryonic physics and the particle nature of dark matter, we find that the extent to which this is possible by altering star formation efficiency is dependent on the specific characteristics of galaxies are being studied. In the context of the anticipated observations of high-redshift galaxies expected to be provided by the JWST, these findings indicate that the JWST will likely be useful in providing better constraints on the baryonic processes of galaxies. However, it is unlikely that the new observations will be able to provide unambiguous answers concerning the particle nature of dark matter.

\section*{Acknowledgements}

D.K. thanks Harvard PRISE for helpful programming and support. S.B. is supported by Harvard University through the ITC Fellowship. S.T. is supported by the Smithsonian Astrophysical Observatory through the CfA Fellowship. 

\section*{Data Availability}
The data underlying this article will be shared on request to the corresponding author.




\bibliographystyle{mnras}
\bibliography{WDM_CDM_JWST} 








\bsp	
\label{lastpage}
\end{document}